\documentclass{optica-article}

\journal{opticajournal} 

\articletype{Research Article}
\usepackage{physics}
\usepackage{lineno}

\begin{document}

\title{Integrated, bright, broadband parametric down-conversion source for quantum metrology and spectroscopy}

\author{René Pollmann,\authormark{*} Franz Roeder, Victor Quiring, Raimund Ricken,Christof Eigner, Benjamin Brecht, and Christine Silberhorn}

\address{
Paderborn University, Department of Physics, Integrated Quantum Optics,\\
Institute for Photonic Quantum Systems (PhoQS),\\
Warburger Straße 100, 33098 Paderborn, Germany
}

\email{\authormark{*}rene.pollmann@upb.de} 


\begin{abstract*} 
Broadband quantum light is a vital resource for quantum metrology and spectroscopy applications such as quantum optical coherence tomography or entangled two photon absorption.
For entangled two photon absorption in particular, very high photon flux combined with high time-frequency entanglement is crucial for observing a signal. 
So far these conditions could be met by using high power lasers driving degenerate, type 0 bulk-crystal spontaneous parametric down conversion (SPDC) sources.
This naturally limits the available wavelength ranges and precludes deterministic splitting of the generated output photons.
In this work we demonstrate an integrated two-colour SPDC source utilising a group-velocity matched lithium niobate waveguide, reaching both exceptional brightness $1.52\cdot10^6\frac{\mathrm{pairs}}{\mathrm{s\,mW\,GHz}}$ and large bandwidth ($7.8\,$THz FWHM) while pumped with a few mW of continuous wave (CW) laser light.
By converting a narrow band pump to broadband pulses the created photon pairs show correlation times of $\Delta \tau \approx 120\,\text{fs}$ while maintaining the narrow bandwidth $\Delta \omega_p \ll  1\,\text{MHz}$ of the CW pump light, yielding strong time-frequency entanglement.
Furthermore our process can be adapted to a wide range of central wavelengths.

\end{abstract*}

\section{Introduction}
As interest in quantum metrology schemes \cite{Mukamel2020} constantly increases, so does the need for tailored quantum light sources. Naturally, the requirements on the quantum state of light and therefore on the source differ, depending on the proposed protocol. 
Therefore, we aim to maximise bandwidth, brightness and time-frequency entanglement \textemdash important benchmarks for entangled two photon absorption \cite{Parzuchowski2021, Dayan2004, Tabakaev2022, gäbler2023}, undetected photon spectroscopy \cite{Lemos2014}, optical coherence tomography \cite{Teich2012} and SU(1,1) interferometry \cite{Lindner2021, Roeder2023}, to name but a few. \\
Even though several schemes enable sensing at incredibly low light intensities down to the single photon level, bright and highly efficient light sources are always advantageous. They yield increased signal to noise ratios, decreased integration times, and simplify the overall setup since higher losses may be acceptable. 
Increasing the source efficiency can also eliminate the need for complex and expensive high power laser systems and simultaneously lessen the need for pump suppression.\\
In this work we demonstrate a tailored parametric down-conversion (PDC) source of high bandwidth driven by a simple continuous wave (CW) diode laser with only a few mW of optical power, whose pair generation rates compare favourably with bulk sources pumped with several Watts of optical power \cite{Tabakaev2022}.
This increase in brightness in terms of the generated pair rate over pump power per bandwidth, comes through the use of integrated optics. 
As the light is confined by the waveguide (WG), the nonlinear interaction length $L$ is not limited by the Rayleigh range of the pump beam. 
Therefore, the brightness of an integrated PDC source scales with $L^2$ in contrast to the linear $L$-scaling of a bulk crystal source \cite{Boyd2003, Regener1988,Fiorentino2007}. 
However, the bandwidth of the phase matching and therefore the generated light, is inversely proportional to the interaction length, generally rendering high brightness and large bandwidth mutually exclusive. 
Here, we show how to overcome this limitation by utilising dispersion engineering \cite{URen2005}, thus achieving large bandwidth and brightness in a two color source. 
Furthermore integrated sources are a significant step towards full integration of the measurement apparatus, with all the inherent advantages of integrated optics such as: miniaturisation, scalability, and the unrivaled stability of a monolithic chip \cite{Tanzilli2012, Pelucchi2022}. 
\section{Concept}
\begin{figure}[ht!]
    \centering
    \includegraphics[width = 0.8 \textwidth]{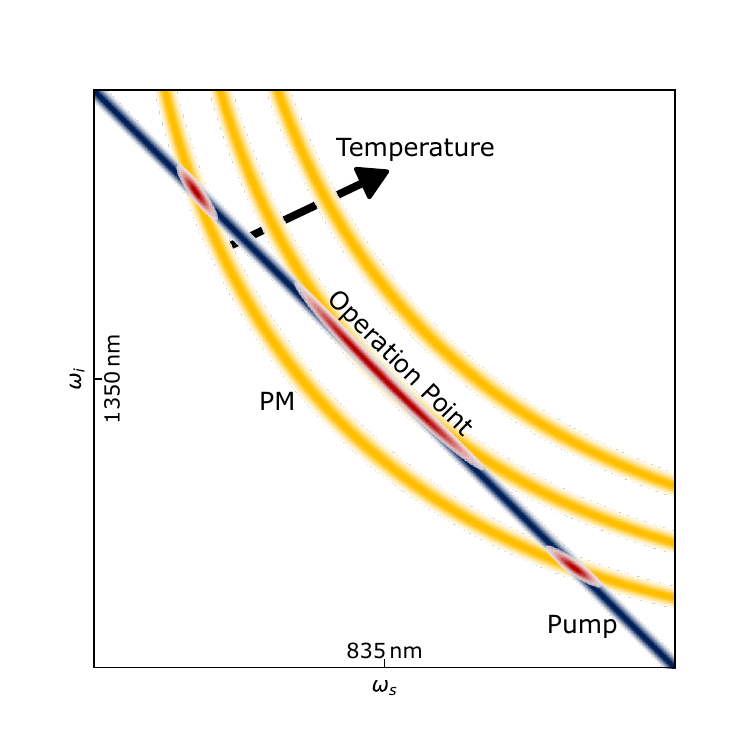}
    \caption{As temperature is increased the PM $\phi$ (yellow) is shifted with respect to the pump $\alpha$, resulting in a broad overlap (red) at the ideal operation point. To low temperatures result in two narrow regions, while to high temperatures result in no overlap at all.}
    \label{fig:schematic}
\end{figure}

PDC is a nonlinear optical process during which one photon of a strong pump field is converted into two daughter photons under conservation of energy and momentum. The generated photons are traditionally called signal ($s$) and idler ($i$). 
The joined spectral amplitude (JSA) $f(\omega_s, \omega_i)$ of the created photons is given by the product of pump and phase matching (PM) functions, $\alpha$ and $\phi$, respectively \cite{Boyd2003, Grice1997}:
\begin{equation}
    f(\omega_s, \omega_i) = \alpha(\omega_p) \cdot \phi(\beta_p,\beta_s,\beta_i),    
\end{equation}
where $\omega_p=\omega_s+\omega_i$ is the pump frequency and the $\beta_{p,s,i}$ are the propagation constants of the involved fields. 
The pump function $\alpha(\omega_p)$ is defined by the properties of the pump laser. 
Due to energy conservation, it always yields anti-correlations in frequency space with a cross-section that is given by the spectrum of the pump laser; in the case of a continuous wave laser, the pump function is to good approximation an anti-diagonal line (blue in fig. \ref{fig:schematic}). 
The PM function represents conservation of momentum (yellow in fig. \ref{fig:schematic}). 
While the bandwidth of the PM is primarily given by the interaction length $L$, its position and shape in frequency space are defined by the propagation constants. 
These are frequency and temperature $T$ dependent material properties $\beta(\omega) = n(\omega,T)\omega/c$ with $n(\omega,T)$ the effective refractive index of the waveguide. 
Therefore, the output of the PDC process can be engineered by appropriate choice and manipulation \cite{mitschke2016fiber, Xin2022} of these parameters. 
As shown in fig. \ref{fig:schematic}, the JSA (red) differs from zero only in the regions where pump and PM overlap.

We aim to produce a broadband PDC output from a narrowband pump, without resorting to short interaction lengths; thus, the PM has to be parallel to the pump (c.f. fig. \ref{fig:schematic}). 
This can can be achieved by matching the group velocities of signal and idler. The PM function of a co linear SPDC source can be written as \cite{Grice1997}:
\begin{equation}
    \phi ( \omega_s, \omega_i ) = \text{sinc} \left( \frac{L}{2} \Delta \beta(\omega_s, \omega_i)  \right) \exp \left( \frac{\imath L}{2} \Delta \beta(\omega_s, \omega_i)  \right). \label{eq:Spectrum}
\end{equation}
Here, 
\begin{equation}
    \Delta \beta (\omega_p, \omega_s, \omega_i) = \beta_p (\omega_p) - \beta_s (\omega_s) - \beta_i (\omega_i),
\end{equation}
is the phase mismatch of the three fields.

The efficiency of the PDC process $\eta = C^2 \frac{L^2}{A} P$ scales linearly with the pump power $P$, and a material and process dependent constant $C^2$. Importantly though, it scales quadratically with $L$ in WGs, as the cross sectional area $A$ of the interaction volume is constant over the full WG length \cite{Tanzilli2012, Boyd2003, Regener1988, Fiorentino2007}. 
The bandwidth of the PM, however, is inversely proportional to $L$. In general this leads to a dilemma: a highly efficient process requires a long interaction length, which in turn limits the bandwidth of the generated PDC.

\begin{figure}[ht!]
    \centering
    \includegraphics[width = 0.95 \textwidth]{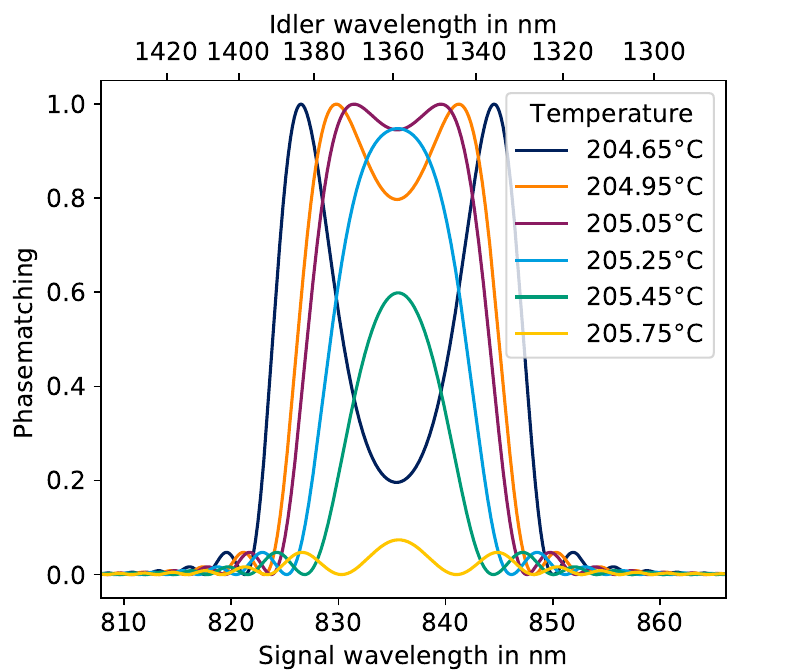}
    \caption{Modelled output spectra  at different WG temperatures.\\
    Variation of the temperature shifts the PM relative to the constant pump ($\lambda_p = 517.4\,$nm), varying the overlap and therefor the spectral shape of the output.}
    \label{fig:theo_spectra}
\end{figure}

To overcome this limitation, the frequency dependence of the phase mismatch has to be minimised over a broad range of frequency combinations which satisfy conservation of energy. 
We rewrite $\omega_p = \omega_s +\omega_i$ and expand $\Delta\beta(\omega_s,\omega_i)$ around the central frequencies $\bar{\omega}_{s,i}$, \cite{URen2005}
\begin{align}
    \Delta \beta (\omega_s, \omega_i) &\approx \eval{\beta_p}_{\bar{\omega}_s + \bar{\omega}_i} - \eval{\beta_s}_{\bar{\omega}_s} - \eval{\beta_i}_{\bar{\omega}_i} - \frac{2 \pi}{\Lambda} \nonumber \\
    &+ \eval{\pdv{\beta_p}{\omega}}_{\bar{\omega}_s + \bar{\omega}_i} (\omega - (\bar{\omega}_s + \bar{\omega}_i))
    -  \eval{\pdv{\beta_s}{\omega}}_{\bar{\omega}_s} (\omega - \bar{\omega}_s)
    -  \eval{\pdv{\beta_i}{\omega}}_{\bar{\omega}_i} (\omega - \bar{\omega}_i). \label{eq:taylor_delta_beta}
\end{align}
As the first line is generally different from zero at the desired frequencies, the periodic poling period $\Lambda$ is introduced to achieve quasi PM \cite{Boyd2003}. 
This shifts the PM towards the desired central frequencies, such that PM and pump function overlap, guaranteeing efficient PDC. 
Note that this positioning of the process in frequency space is highly temperature dependent and can therefore be fine tuned in the experiment by choosing an appropriate operating temperature, see figs \ref{fig:schematic}, \ref{fig:theo_spectra} and \ref{fig:measured_spectra}. 
The derivatives in the second line in \eqref{eq:taylor_delta_beta} are the inverse group velocities of the fields in the WG, describing the angle of the PM in frequency space. 
As the group velocities are less temperature dependent, the angle of the PM dose not change significantly with temperature.
Since we assume the pump to be monochromatic the frequency mismatches can be rewritten as $\Delta\Omega = (\omega - \bar{\omega}_s) = - (\omega - \bar{\omega}_i)$ utilising conservation of energy, thus the term containing the pump vanishes, since $\omega - (\bar{\omega}_s + \bar{\omega}_i) = \omega - \bar{\omega}_s + \omega - \bar{\omega}_i = \Delta\Omega - \Delta\Omega = 0$ and \eqref{eq:taylor_delta_beta} reduces to
\begin{equation}
    \Delta \beta (\omega_s, \omega_i) \approx  \frac{1}{v_{g,s}} \Delta\Omega - \frac{1}{v_{g,i}} \Delta\Omega. \label{eq:GVM_condition}
\end{equation}

Therefore, matching the group velocities of signal and idler $v_{g,s} = v_{g,i}$ minimises $\Delta \beta$ over a broad  range of frequency detunings $\Delta\Omega$, leading to a broadband output state originating from a narrow band input, independent of the interaction length $L$ \cite{URen2005, Vanselow2019}. 
Note that group-velocity matching is trivially guaranteed for degenerate type 0/I processes, where both photons are indistinguishable; these processes, however, do not satisfy the fundamental requirements of quantum-enhanced sensing protocols as sensing with undetected photons for example requires degenerate sources \cite{Lemos2014, Teich2012, Lindner2021, Roeder2023}. 
In a type II process, where the generated photons have orthogonal polarization on the other hand, the birefringence of the material can be exploited to achieve group velocity matching. 
Then, one can always identify a set of $\{\omega_p, \omega_s, \omega_i\}$ given one of the frequencies for which $v_{g,s}=v_{g,i}$ holds.\\
With $\{\omega_p, \omega_s, \omega_i\}$ known, we can calculate the poling period $\Lambda$ required for quasi-phase matching and fabricate the source.
We modeled the expected output spectra of our source, see fig. \ref{fig:theo_spectra}. 
Here, one can immediately observe the strong temperature dependence of the process, discussed in detail in \ref{cha:spectra}. 
This can be understood in the following way: group-velocity matching ensures that the PM is aligned parallel to the pump function in the $(\omega_s,\omega_i)$ plane.
It does not, however, guarantee that the centers of the pump and PM functions overlap optimally.
This overlap can be achieved by tuning either the pump wavelength, thereby moving the pump function, or by adjusting the sample temperature, thereby moving the PM function.
For reasons of experimental convenience, we chose to adjust the sample temperature while keeping the pump fixed, c.f. fig. \ref{fig:schematic}.
If the temperature of the WG is too low, pump and PM intersect in two separate spectral regions, leading to two peaks in the output spectrum, as shown in dark blue in  fig. \ref{fig:theo_spectra}. 
An increasing sample temperature leads to a gradual merging of the two peaks, until an optimum operation point is reached, purple in fig. \ref{fig:theo_spectra}. A further increase of the temperature results in a loss of overlap between the pump and PM and consequently a vanishing output, yellow in fig. \ref{fig:theo_spectra}. 
Note, that the direction and magnitude of this shift is dependent on the material, type of process and central wavelengths. 
Due to fabrication imperfections and inaccuracies of the underlying WG and material models, the ideal working point for any given WG has to be found experimentally by measuring the output spectrum at different temperatures, see fig. \ref{fig:measured_spectra} for the measured spectra.




\section{Measurements}
We fabricated a WG source designed to produce group velocity matched type II SPDC driven by a $517.4\,$nm CW pump laser.
To validate our theoretical model we analyse the temperature dependency of the output spectra and determine the ideal operation point (see section \ref{cha:spectra}).
Then, we measure the brightness of our source and compare it with existing bulk-crystal and integrated sources in section \ref{cha:brightnes}.

\subsection{Integrated PDC source and setup}\label{cha:setup}
\begin{figure}[ht!]
    \centering
    \includegraphics[width = 0.8 \textwidth]{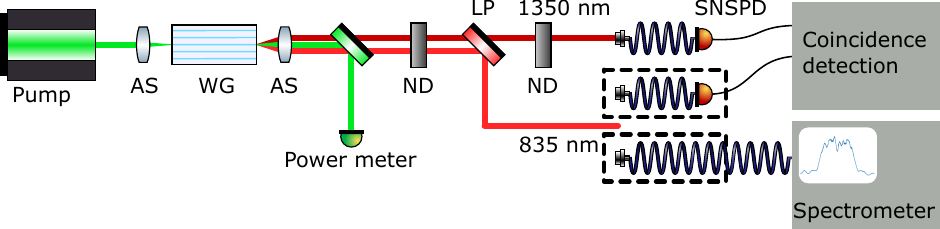}
    \caption{Experimental setup used for measuring source brightness (upper box) and PDC spectra (lower box), respectively. 
    The pump laser is coupled to the waveguide (WG) by an aspheric lens (AS) and generated light is then collimated by another AS. 
    The pump light is reflected by a interference filter and sent to a power meter. 
    Signal and idler are split up by a long pass filter (LP) and coupled to single mode fibers. 
    For measuring source brightness, both fields are sent to single photon detectors. 
    To avoid saturating the detectors neutral density filters (ND) are placed in the beam paths as needed. 
    For measuring PDC spectra, the signal beam is sent to a spectrometer, while the idler beam is discarded.}
    \label{fig:measurment_setup}
\end{figure}

We employ a 40\,mm long, periodically poled Ti-indiffused WG in z-cut LiNbO$_3$ with a poling period of $\Lambda = 3.69\,\upmu$m, which we fabricated in house. 
To suppress Fabry-Perot oscillations in the WG and increase overall efficiency, the output facet is anti-reflection coated for all three wavelengths. 
The LiNbO$_3$ sample is placed in a custom build WG oven driven by a PID controller (Oxford mercury). 
To further improve temperature stability the WG oven is housed in two nested boxes with holes cut for the beams to pass. 
A scheme of the experimental setup is shown in fig. \ref{fig:measurment_setup}. 
The source is pumped with a grating stabilised CW diode laser (Toptica DL-Pro) at $\lambda_p = 517.4\,$mn, which is coupled into the WG via an aspheric lens (Thorlabs A110TM). 
The output is subsequently collimated via another asperic lens (Thorlabs C397TMD) and the pump field is separated by a notch filter (Semrock  NF03-532E-25 angle tuned to 517.4\,nm) and sent to a power meter (Thorlabs S132C). 
Residual pump light is furhter suppressed by an additional filter (Thorlabs FEL800) before signal and idler fields are split on a dicroic beamsplitter (Semrock LP980) and coupled into single mode fibers (780HP and SMF28) with asperic lenses (Thorlabs A240TM-B and A110TM-C) to interface with the detectors. 
Here, we employ single mode fibres as spatial mode filters, allowing the selection of one single spatial mode of the WG, therefore suppressing signals from competing processes such as fluorescence, black body radiation and ambient light.

\subsection{Spectra and bandwidths}\label{cha:spectra}
\begin{figure}[ht!]
    \centering
    \includegraphics[width = 0.95 \textwidth]{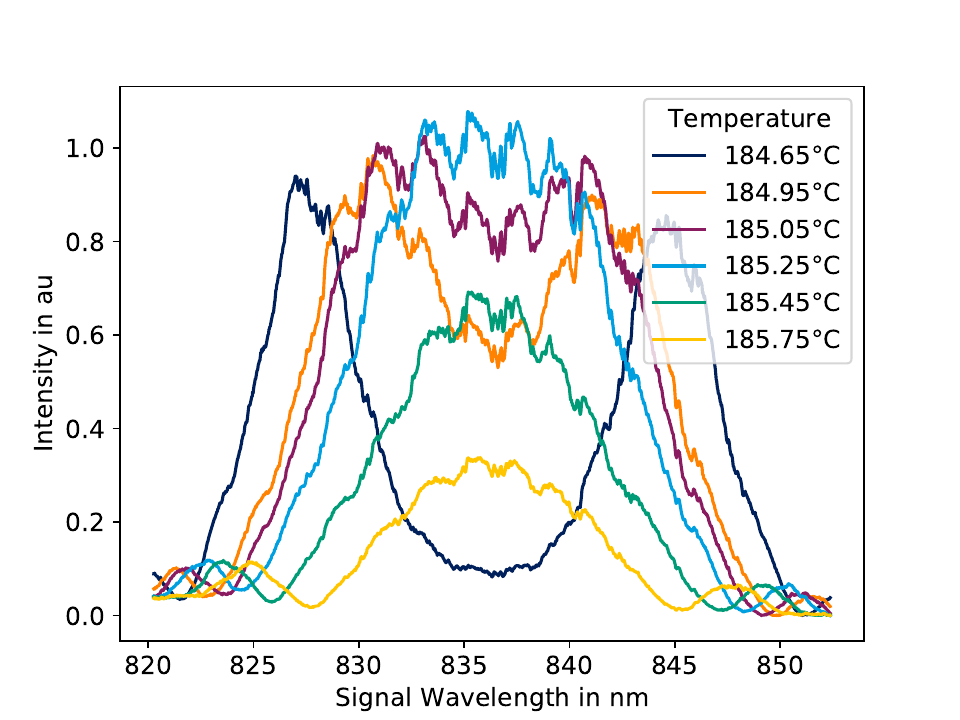}
    \caption{Measured signal spectra at different WG temperatures. By increasing the temperature past the ideal operation point (purple) the bandwidth can be reduced (light blue), before source brightness drops due to a reduced overlap between pump and PM (green and yellow).}
    \label{fig:measured_spectra}
\end{figure}

\begin{figure}[ht!]
    \centering
    \includegraphics[width = 0.95 \textwidth]{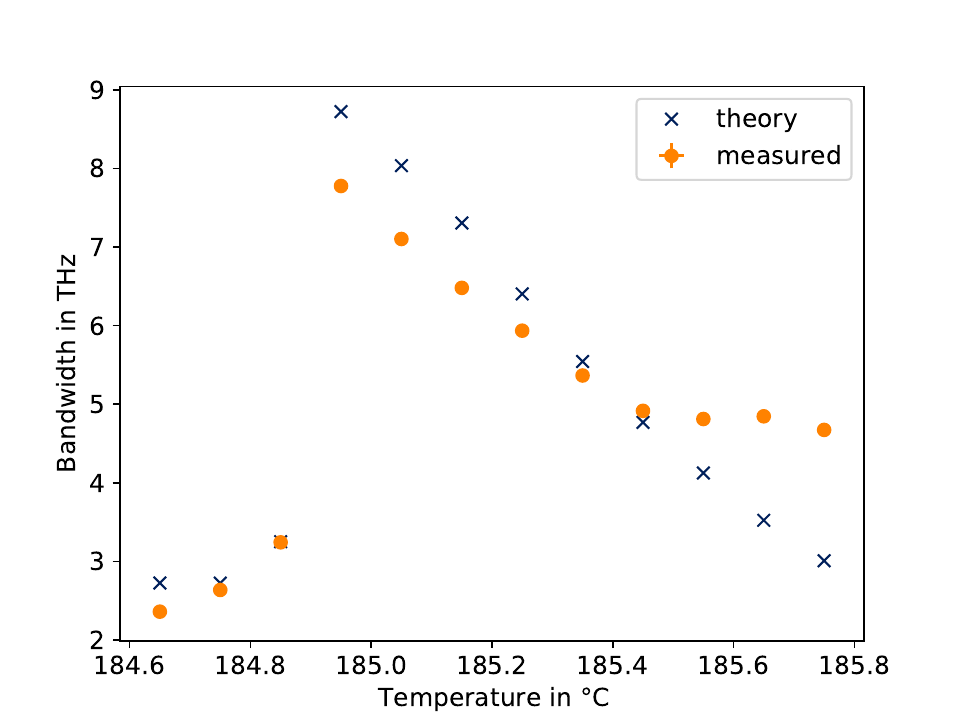}
    \caption{Measured and simulated bandwidth. Note that the theory is shifted by $-20\,^\circ$C to overlay with measured data. Error bars are smaller than the size of the markers. For more information, see the text.}
    \label{fig:measured_bandwidth}
\end{figure}

As discussed above, the output spectrum is highly temperature dependent. Therefore the signal spectrum was measured at different WG temperatures with a single photon sensitive grating spectrometer with a resolution of 30\,GHz (Andor Shamrock SR-500i spectrograph with Newton 970P EMCCD - camera) and plotted in fig \ref{fig:measured_spectra}. 
All spectra captured with this particular spectrometer show small oscillations with a $\approx2.1$\,nm period. 
As these stay fixed in amplitude and position over all measurements taken with this device we attribute them to Fabry-Perot oscillations in the CCD cover glass.
As expected from the model shown in figs. \ref{fig:schematic} and \ref{fig:theo_spectra}, the spectrum shows two distinct peaks at low temperatures (dark blue), originating from the two intersections of pump and PM function. 
As the temperature is increased, these peaks gradually move towards each other and finally fuse into one broad peak, yielding maximal bandwidth of 7.8\,THz (purple). 
A further increase in temperature decreases the overlap of pump and PM and therefor decreases both bandwidth an efficiency of the process (yellow).
Comparing the measured spectra shown in fig. \ref{fig:measured_spectra} with the simulated ones shown in fig. \ref{fig:theo_spectra}, we find the overall shape to be reproduced very well, albeit the temperatures in the simulation are 20\,$^\circ$C higher and the decrease in bandwidth (BW) and efficiency at high temperatures is more pronounced.
We attribute this temperature offset to the lack of calibration of the temperature sensor used in the WG oven as well as inaccuracies in the pump wavelength and the temperature dependence of our WG model which is beyond the scope of this work \cite{Santandrea2019_1, Santandrea2019_2, Santandrea2019_3}. 
To further investigate the differences between simulation and measurement, we plot the calculated and measured BW's, defined as full width at half maximum (FWHM), over the WG temperature in fig. \ref{fig:measured_bandwidth}. For comparability, we displace the simulation results by -20\,$^\circ$C. 
As expected, both data sets follow the same general trend. 
The BW slowly increases with rising temperature, as the angle between the pump and PM at the two intersections gradually decreases and the two peaks approach each other. 
We expect the experimental PM to be broadened by inhomogeneities in the WG as well as the in periodic poling \cite{Santandrea2019_1, Santandrea2019_2, Santandrea2019_3}. Consequently, we run the simulation with an effective length of $L_{model} = 25\,$mm, artificially broadening the PM to better match the measured data. 
All further simulation parameters are matched to the experimental parameters listed in section \ref{cha:setup} as closely as possible.\\
At 184.9$\,^\circ$C we observe a steep increase in BW as the two peaks merge\footnote{We consider the peaks merged if the local minimum in between them exceeds 50\% of the peaks maximum value, consistent with the FWHM.} and the ideal opreation point is reached. 
While further heating decreases the BW in model and measurement, in the theoretical model the BW decreases quicker than in the experiment. This is likely caused by the same inaccuracies of the WG model that cause the -20\,$^\circ$C shift of the operation point. 
At higher temperatures this trend reverses and the measured BW becomes larger. We attribute this to imperfections in the periodic poling, leading to spectral artefacts in the PM, which are now amplified considerably by the displaced pump.

\subsection{Time-frequency entanglement}
We estimate a minimal correlation time that is, the simultaneity of the generated signal and idler photons, of $\Delta \tau = 120$\,fs (FWHM) by Fourier transforming the measured signal spectrum (purple in fig, \ref{fig:measured_spectra}) while assuming a flat spectral phase.
At the same time, the spectral bandwidth of the bi-photon is given by the bandwidth of the pump, which is on the order of $\Delta \omega_p \lesssim 1$\,MHz for the free running grating stabilised diode laser \cite{laser_data_sheet} utilised in the experiment. 
With this the time bandwidth product of the bi-photon evaluates to:
\begin{align}
    \Delta \tau \cdot \Delta \omega_p \approx 120\cdot10^{-15}\,\text{s} \cdot 1\cdot10^{6}\,\text{Hz} = 1.20\cdot10^{-7} \ll 0.93,
\end{align}
where we calculate the classical Fourier limit of $\Delta\omega\cdot\Delta\tau=0.93$ from the Fourier transform of the PDC spectra.
The bi-photon time-bandwidth product beats the classical limit by at least seven orders of magnitude, which demonstrates strong time-frequency entanglement between the generated photons \cite{Brecht2013}.

\subsection{Brightness}\label{cha:brightnes}
\begin{figure}[ht!]
    \centering
    \includegraphics[width = 0.95 \textwidth]{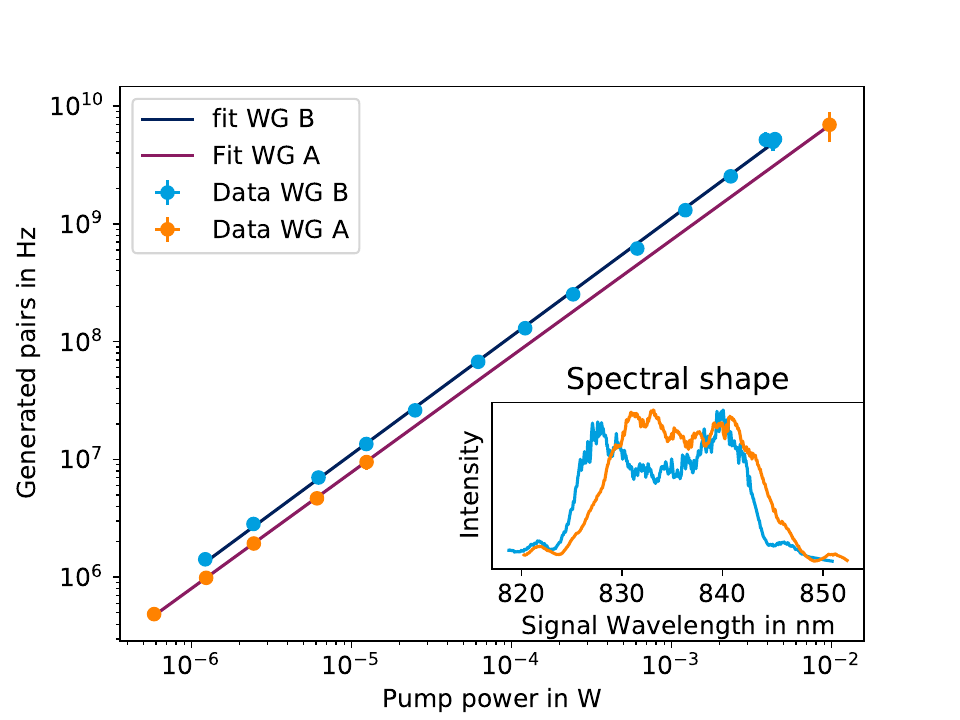}
    \caption{Measured generated pair rate (see eq\eqref{eq:gen_pairs}) over the transmitted pump power for two WGs. While WG A shows higher maximal pair rates, WG B shows a slightly higher efficiency. During the measurements we took special care to keep the PDC spectra, shown in the inset, constant. Error bars correspond to shot noise and the accuracy of the power meter in use.}
    \label{fig:measured_brightness}
\end{figure}

For measuring the source brightness, we detect the signal and idler photons with superconducting nanaowire single photon detectors (SNSPD) (Single Quantum EOS) optimesed for 800\,nm and 1550\,nm, respectively (fig \ref{fig:measurment_setup}) and analyse the measured click pattern with a time to digital converter (Swabian Instruments Time-Tagger 20). 
We chose a coincidence time window of $t_\text{w} = 1400$\,ps, which optimizes coincidence events while optimally suppressing accidental events.
The signal beam also passes an additional 50\,nm wide bandpass filter centred at 825\,nm (Edmund Optics \#86-957) to suppress spurious fluorescence centered around 890\,nm. 
As the source is easily bright enough to saturate the detectors, a selection of ND-filters (Thorlabs NEK01) is used to attenuate the fields in front of the fibre coupling.
The attenuation was chosen to achieve single count rates of $\approx300\,$kHz in either arm, striking a balance between low coincidence rates and events lost to detector dead times. 
We had to place additional filters in the idler arm because of their reduced attenuation at idler wavelengths.
We obtained the overall transmission in two ways: On the one hand, we directly measured the transmission using classical laser light. On the other hand, we deduced the efficiency from measuring coincidence and single count rates and calculated the associated Klyshko efficiencies \cite{Klyshko1980}.
Since the WG defines the spatial mode of signal and idler, we could easily measure fibre coupling efficiencies and found excellent agreement between measured losses and Klyshko efficiencies.
We subtracted the accidental coincidence rate $r_{cA}$ from the measured coincidence count rate and divided by the Klyshko efficiencies $\eta_{s,i}$ to obtain the generated pair rate

\begin{align}
    r_{c} = \frac{r_{cM} - r_{cA}}{\eta_s \cdot \eta_i} = \frac{r_{sM} \cdot r_{iM}}{r_{cM} - r_{cA}}. \label{eq:gen_pairs}
\end{align}
Here, $r_{cA} = 2 r_s r_i t_\text{w}$ and the $r_{sM,iM}$ are the measured single count rates \cite{Tanzilli2001}.\\
In fig. \ref{fig:measured_brightness}, we plot the generated pair rate as function of the transmitted pump power after the WG.
We measured two WGs, with WG A\footnote{also used for spectral characterisation} heated to 185.05\,$^\circ$C and WG B to $199.75\,^\circ$C yielding a bandwidth of 7.8\,THz and 7.5\,THz, respectively. 
Special care was taken to keep the PDC spectrum, shown in the inset, constant during all measurements. 
WG A shows a smaller conversion efficiency of $(0.2754 \pm 0.00013)\,\upmu$W/W or $(1.14 \pm 0.07 )\cdot 10 ^{12}$\,Pairs/s/W compared to WG B at $(0.4793 \pm 0.00009)\,\upmu$W/W and $(0.66 \pm 0.02)\cdot 10 ^{12}$\,Pairs/s/W respectively.
Due to higher achievable pump powers we find a larger maximal generated pair rate of $(7 \pm 2)\cdot 10 ^{9}$\,Hz in WG A compared to $(5.1 \pm 0.8)\cdot 10 ^{9}$\,Hz observed in WG B.
These rates were limited by inefficient coupling of the pump beam to the WGs due to Fresnel reflections on the uncoated surfaces and improper mode matching, as the elliptical beam profile of the pump laser was not corrected for. \\
High optical power densities can cause random modulations of the refractive index in LiNbO$_3$ via the photorefractive effect, which are especially detrimental in integrated nonlinear devices. 
However, it is well known, that the photorefractive effect can be suppressed by operating the device at sufficiently high temperatures \cite{Rams2000}. 
It is worth mentioning that our maximum rates were limited by available pump laser power and not by photorefraction.

A comparison of similar photon pair sources is shown in Table \ref{tab:comparisson}, where we limit ourselves to two materials, namely periodically poled LiNbO$_3$ (LN) and periodically poled potassium titanyl phosphate (KTP). 
We compare type 0 and type II processes regarding reported bandwidths and brightness, which is defined as generated pair rate over pump power per bandwidth. This enables a comparison between narrow and broad bandwidth sources.\\
Our source shows comparable bandwidth to degenerate type 0 sources, a result of the dispersion engineering.
In a spectrally degenerate type 0 PDC, signal and idler photons are indistinguishable by definition, hence exhibit the same group velocity.
Therefore, such sources necessarily feature large spectral bandwidth.
However, signal and idler having the same central wavelength makes these sources unsuitable for some metrology applications \cite{Lemos2014, Teich2012, Lindner2021, Roeder2023}. 
Our source's bandwidth is only surpassed by bulk-crystal sources also utilising GVM \cite{Lindner2021, Vanselow2019}. 
As these sources shift the idler wavelength to the mid infrared, they can also minimise group velocity dispersion and therefore reach these extreme bandwidths.
However, this makes it difficult to detect the idler photons directly, prohibiting actual photon counting and thus determination of the source brightness. 
Nonetheless, the source shown in \cite{Vanselow2019} is the only other broadband type II source reported in literature to the best of our knowledge.
We emphasize that our design principles can readily be extended to other wavelengths and therefore the operation at NIR wavelengths is not a fundamental limitation of our source.

In terms of brightness WG sources outperform bulk-crystal sources by several orders of magnitude \cite{Tanzilli2001, Fiorentino2007}, showing the advantage gained by tight confinement and long interaction length.
When comparing type 0 and type II processes, one has to take the reduced magnitude of the second order nonlinear coefficient into account. 
For LN this accounts for a $\approx 36$-fold advantage in brightness for type 0 processes. This can be overcompensated by utilising WGs, as can be seen when comparing the source reported here to the source shown in \cite{Tabakaev2022}.

When compared to other type II sources we see our source reaching the same brightness as narrow band sources. 
The lower brightness when compared to the WG shown in \cite{Fiorentino2007} can be explained by the tighter confinement in the much smaller KTP WG as well as the larger field overlap of a degenerate process compared to a non degenerate, while the difference between the second order nonlinear coefficients of LN and KTP is negligible.



While our source dose not outperform other sources in bandwidth or brightness, it simultaneously achieves competitive results in both metrics while keeping the output distinguishabel, therefore combining three traditionally pairwise exclusive parameters at once.

\begin{table}
    \begin{tabular}{|c|c|c|c|c|}
    \hline
        Source type                     & Material  & method        & bandwidth     & brightness in  \\ 
                                        &           &               & in THz        & $\dfrac{\text{pairs}}{\text{s\,mW\,GHz}} $\\ \hline
        Type 0 \cite{Tabakaev2022}      & LN        & deg. Type 0   & 7             & $2.45\cdot 10^4$ \\   \hline
        Type 0 \cite{Lindner2021}       & LN        & GVM           & 21            & $<6.99\cdot 10^3$\\ \hline
        Type 0 \cite{Dayan2005}         & KTP       & deg. Type 0   & 8.2           & $3.27\cdot 10^4$ \\ \hline 
        Type 0 \cite{Tanzilli2001}      & WG LN     & deg. Type 0   & 5.2           & $1.27\cdot 10^6$\\ \hline
        Type 0 \cite{Kaiser2016}        & WG LN     & deg. Type 0   & 10            & $1.20\cdot 10^6$ \\ \hline
        Type 0 \cite{gäbler2023}        & WG LN     & deg. Type 0   & $\approx9$    & $6.34\cdot 10^4$\\ \hline\hline
        Type II \cite{Vanselow2019}      & KTP       & GVM           & 25            & NA\\ \hline
        Type II \cite{Fiorentino2007}    & KTP       &               & 0.3           & $6.00\cdot 10^3$\\ \hline
        Type II \cite{Fiorentino2007}    & WG KTP    &               & 0.3           & $2.43\cdot 10^5$\\ \hline
        Type II \cite{Harder2013}        & WG KTP    &               & 0.496         & $2.43\cdot 10^5$\\ \hline
        Type II \cite{Montaut2017}       & WG LN     &               & 0.22          & $1.13\cdot 10^5$\\ \hline \hline
        Type II WG A [this work]         & WG LN     & GVM           & 7.5           & $1.52\cdot 10^5$\\ \hline
        Type II WG B [this work]         & WG LN     & GVM           & 7.8           & $8.49\cdot 10^4$\\ \hline
    \end{tabular} 
    \caption{Comparison of broadband and type II sources.\\
    Type 0 sources show large bandwidth sacrificing distinguishability, while type II sources only reach narrow bandwidth without utilising GVM. As discussed above, waveguided sources consistently outperform bulk sources in terms of brightness.}
    \label{tab:comparisson}
\end{table}

\section{Conclusion}
In this work we have introduced a dispersion-engineered, integrated source of ultra-broadband parametric down-conversion. In contrast to other sources that are based on spectrally degenerate type 0 processes, our source employs a type II process with distinguishable signal and idler. 
A careful matching of the group velocities of the generated photons then facilitates a broad spectral bandwidth for highly non-degenerate photon frequencies. 
Our source overcomes the inherent trade off between brightness and bandwidth that one typically faces in integrated quantum photonics applications.\\
Our source features a spectral bandwidth of $7.8\,$THz together with an exceptional brightness of $1.52\cdot10^6\frac{\mathrm{pairs}}{\mathrm{s\,mW\,GHz}}$. 
These numbers are on par with non-engineered sources based on type 0 PDC in bulk crystal, showcasing the efficiency gain through integration. 
The non-degeneracy of the wavelengths renders our source ideally suited for quantum spectroscopy applications that require a high degree of control. 
Prime examples are entangled two-photon absorption and SU(1,1) interferometry, as well as applications based on undetected photons.\\
The underlying design principle of our source can be adapted to provide emission at user-chosen target wavelengths. 
Hence, we expect this source to become a useful tool for future quantum spectroscopy applications, which will benefit from the increased stability and robustness offered by integrated optics.
\begin{backmatter}

\bmsection{Funding}
We acknowledge financial support from the Federal Ministry of Education and Research (BMBF) via the grant agreement no. 13N15065 (MiLiQuant) and no. 13N16352 (E2TPA)
\bmsection{Disclosures}
The authors declare no conflicts of interest.
\bmsection{Data availability}
Data underlying the results presented in this paper are not publicly available at this time but may
be obtained from the authors upon reasonable request.
\end{backmatter}

\bibliography{sources}

\end{document}